\documentclass[showpacs,preprint,amssymb]{revtex4}
\usepackage{graphicx}% Include Figure files
\usepackage{dcolumn}% Align table columns on decimal point
\usepackage{bm}% bold math
\def\be{\begin{equation}} 
\def\ee{\end{equation}}
\def\bea{\begin{eqnarray}} 
\def\eea{\end{eqnarray}}
\def\line{\hbox to \hsize}    
\def\frac #1#2{{#1\over #2}}

\def\1{\mbox{\bf 1}}

\begin{document}
%\draft %(only for revtex) 

\title{The classical hydrodynamics of the  Calogero-Sutherland model}

\author{ MICHAEL STONE}

\affiliation{University of Illinois, Department of Physics\\ 1110 W. Green St.\\
Urbana, IL 61801 USA\\E-mail: m-stone5@uiuc.edu}   

\author{INAKI ANDUAGA}

\affiliation{University of Illinois, Department of Physics\\ 1110 W. Green St.\\
Urbana, IL 61801 USA\\E-mail: anduaga2@uiuc.edu}   

\author{LEI XING} 

\affiliation{University of Illinois, Department of Physics\\ 1110 W. Green St.\\
Urbana, IL 61801 USA\\E-mail: leixing2@uiuc.edu}   

\begin{abstract}  

We explore  the classical version of the mapping, due  to Abanov and Wiegmann,  of  Calogero-Sutherland hydrodynamics  onto  the Benjamin-Ono equation ``on the double.''  We illustrate the mapping by constructing the soliton solutions to the hydrodynamic equations, and show how certain  subtleties   arise from the need to include corrections to the na{\"\i}ve
replacement of singular sums by principal-part integrals.

\end{abstract}


\maketitle

\section{Introduction}

The Calogero-Sutherland  family of   models \cite{calogero,calogero-sutherland} consist of point particles moving on a line or circle and interacting with a  repulsive inverse-square potential.   Both the  classical and quantum versions are completely integrable, and are the subject of an extensive literature \cite{CSreviews}.  The models have application to one-dimensional electron systems \cite{gutman}, 
and to the two-dimensional quantum Hall effect \cite{polychronakos_chern}.      

It is possible to consider a hydrodynamic   limit of the Calogero-Sutherland models in which the distribution  and velocity   of the particles are described by continuous fields $\rho(x)$ and $v(x)$ respectively.  The equations of motion of these fields  possess  solitary wave  solutions, and also periodic   solutions that interpolate between small amplitude sound waves and large amplitude   trains of solitary waves    \cite{polychronakos_soliton,jonke}. 
The quantum  version of the  hydrodynamics  provides an extension of the usual theory of bosonization of relativistic  ({\it i.e.}\ linear dispersion) electron systems  to systems  where band-curvature effects become important \cite{glazman}. Because of the  singular nature of the Calogero-Sutherland interaction,  the hydrodynamic  limit is rather more subtle  than one might expect.  Considerable insight into this limit  has been  provided  by Abanov and Wiegmann who have shown \cite{abanov} that it is equivalent to a ``doubled" version of     Benjamin-Ono  dynamics   \cite{benjamin,ono}.   The  Benjamin-Ono equation, a member of an infinite  hierarchy of integrable partial differential equations,  was originally introduced to describe waves in stratified fluids.  The connection has lead  to a number of novel predictions  for the evolution of one-dimensional electron gasses \cite{bettelheim}. 

In this paper we will apply the tools  of \cite{abanov} to recover the classical soliton solutions found in \cite{polychronakos_soliton,jonke} and, in doing so, illustrate the structure  of the Calogero-Sutherland $\leftrightarrow$ Benjamin-Ono  mapping. A secondary   aim  is  to compare the origin of the hydrodynamic-limit subtleties   in the classical model with their origin in the quantum system.  In the  quantized  model  they arise because of  the need to convert operators that are hermitian with respect to one of the  two natural  inner products on the Calogero-Sutherland Hilbert space into operators that are hermitian with respect to the other \cite{stone_gutman}. 
In the classical model  the subtleties  arise because we need to  include corrections to the na{\"\i}ve limit of singular sums.
 
In section \ref{SEC:pole_ansatz}, we review the connection between
the   Calogo-Sutherland and   Benjamin-Ono equations of motion. In section \ref{SEC:shepherd} we consider an appealing, but  overly na{\"\i}ve, version of the hydrodynamic limit and reveal its failings. In section \ref{SEC:correction} we illustrate  how these shortcomings are eliminated by including the first non-trivial order in an asymptotic expansion for  the velocity field. In section \ref{SEC:double}  we consider the general mapping. An appendix contains  proofs of  some  results used    in the main text.

\section{Calogero-Sutherland from  the Benjamin-Ono pole ansatz}
\label{SEC:pole_ansatz}

The classical Benjamin-Ono  equation  \cite{benjamin,ono} is  the   nonlinear and and non-local  partial differential equation 
\be
\dot u +u\,\partial_x u= \textstyle{ 1\over 2}  \lambda  (\partial^2_{xx}u)_H,
\label{EQ:boequation}
\ee
where    $f_H$ denotes   the  Hilbert transform of $f$:
\be
f_H(x)\, \stackrel{\rm def}{=}\,\frac{\rm P}{\pi}\int_{-\infty}^{\infty} \frac{1}{x-\xi}f(\xi)\,d\xi.
\label{EQ:realHilbert}
\ee
If we introduce a  Poisson bracket
\be
\{u(x),u(x')\}= 2\lambda \pi \, \partial_x \delta(x-x'),
\label{EQ:poisson1}
\ee
and Hamiltonian
\be
H_{\rm BO}= \frac 1{2\lambda \pi} \int_{-\infty}^{\infty}\left\{ \frac 16 u^3 - \frac \lambda 4 u (u_x)_H\right\} dx,
\ee 
then (\ref{EQ:boequation}) can be written   as $\dot u(x,t) =\{H_{\rm BO},u(x,t)\}$.  This Hamiltonian system   possesses  infinitely many Poisson-commuting integrals of   motion \cite{case}.

Following \cite{PLC}, we   seek solutions of   (\ref{EQ:boequation}) as a sum of poles
\be
u(x,t) = \sum_{j=1}^N \frac{i\lambda}{x-a_j(t)} -  \sum_{j=1}^M \frac{i\lambda}{x-b_j(t)}.
\label{EQ:poleansatz}
\ee
The  poles at  $a_j(t)$, $j=1,\ldots,N$,  lie   below the real axis while   the poles at $b_j(t)$, $j=1,\dots, M$,   lie  above it. In \cite{PLC},   the numbers of $a_j$ and $b_j$ poles  were set  equal and   $b_j(t)=a_j^*(t)$.    These conditions were imposed  to ensure that    $u(x,t)$  was   real.  We will not make these   assumptions, so our field $u(x,t)$ is  not necessarily real-valued.

We insert the  ansatz (\ref{EQ:poleansatz}) into   (\ref{EQ:boequation}) and use 
\be
\left( \frac{1}{x-a}\right)_H= \cases{\displaystyle{\frac{-i}{x-a}}, &  ${\rm Im\,} a<0$\smallskip\cr
                                                              \displaystyle{\frac{+i}{x-a}}, &  ${\rm Im\,} a>0$}
\ee
to find
\bea 
&&\left\{\sum_{k=1}^N\frac{i\lambda \dot a_k}{(x-a_k)^2}-\sum_{k=1}^M\frac{i\lambda\dot b_k}{(x-b_k)^2}\right\}
- \left\{ \sum_{j=1}^N \frac{i\lambda }{x-a_j} -  \sum_{j=1}^M\frac{i\lambda}{x-b_j}\right\}   \left\{\sum_{k=1}^N\frac{i\lambda }{(x-a_k)^2}-\sum_{k=1}^M\frac{i\lambda }{(x-b_k)^2}\right\}\nonumber\\
&& \qquad \qquad - \lambda\left\{  \sum_{j=1}^N\frac{\lambda }{(x-a_j)^3}+ \sum_{j=1}^M\frac{\lambda}{(x-b_j)^3}\right\}=0.
\eea
All  terms with $1/(x-a_i)^3$ and $1/(x-b_i)^3$ cancel directly.  The remaining terms can be simplified by exploiting    the identity
\be
\frac{1}{(x-c_1)}\frac{1}{(x-c_2)^2}+\frac{1}{(x-c_2)}\frac{1}{(x-c_1)^2}=\frac{1}{(c_2-c_1)}\frac{1}{(x-c_2)^2}+\frac{1}{(c_1-c_2)}\frac{1}{(x-c_1)^2},
\label{EQ:fact351}
\ee
 to rearrange  them    as a sum of    $1/(x-a_i)^2$'s and $1/(x-b_i)^2$'s with   $x$-independent coeffcients. Now the  set of  $1/(x-a_i)^2$'s and $1/(x-b_i)^2$'s is  linearly independent, and the vanishing of their  individual coefficients requires 
 \footnote{These equations are similar to, but not identical with  Kelvin evolution of point vortices. The difference is that for a set of vortices of strength $\kappa_i$ at point  $z_i$ we have equations of the form
 $$
 \dot z_i^* =\frac {1}{2\pi i} \sum_{j\ne i} \frac{\kappa_j}{z_i-z_j}.
 $$
 in which the  complex conjugate of $\dot z_i$ is given by a pole  sum  involving only  unconjugated $z_i$.}
\bea
i\dot a_j &=& \sum_{k;\, k\ne j} \frac{\lambda}{(a_k-a_j)} - \sum_{k=1}^M \frac{\lambda}{(b_k-a_j)},\
\label{EQ:a_velocity}\\
i\dot b_j &=& \sum_{k;\,  k\ne j} \frac{\lambda }{(b_j-b_k)} - \sum_{k=1}^N \frac{\lambda }{(b_j-a_k)}.
\label{EQ:b_velocity}
 \eea
We have found   $N+M$ evolution equations for  $N+M$ variables, and so the pole ansatz is internally consistent.
  
We  now compute $\ddot a_j$ by differentiating (\ref{EQ:a_velocity}),  and then  using equations 
(\ref{EQ:a_velocity}),  (\ref{EQ:b_velocity}) to eliminate the $\dot a_i$ and $\dot b_i$'s. After some labour involving repeated   use of (\ref{EQ:fact351})  we find that     
 \be
 \ddot a_j = \sum_{k;\,k\ne j} \frac{2\lambda^2}{(a_j-a_k)^3}. 
 \label{EQ:calogeroeq}
 \ee
Remarkably, the $b_j(t)$ do not appear.   The  only role of the $b_j$'s  in the $a_i$-pole dynamics is  that the  complex parameters $b_j(0)$  determine  the  (complex) initial velocities $\dot a_i(0)$ of the  $a_i(t)$-poles. Once these initial conditions are established, the $a_i$ poles evolve autonomously according to (\ref{EQ:calogeroeq}).

The $N$   equations   (\ref{EQ:calogeroeq})  are   a complex version of the Calogero model equations of motion.  As for the real-$a_i$ case,  they can be derived    from the many-body Lagrangian
 \be
 L_{\rm Calogero}= \frac 12 \sum_{i=1}^{N} \dot a_i^2 - \frac 12 \sum_{i,j;\, i\ne j}\frac{\lambda^2}{(a_i-a_j)^2}.
 \label{EQ:calogeroact}
 \ee
 
 We may  site   our initial poles  so that if we place  a pole at $z$, then we  also place one at $z+2\pi n$ for any integer $n$. The resulting  $2\pi$ periodicity will  be preserved by the subsequent evolution. The sums 
 \bea
 \lim_{M\to \infty} \left\{\sum_{n=-M}^M \frac 1{z+ 2\pi n}\right\}&=& \frac 12 \cot\left(\frac z 2\right),\\
 \sum_{n=-\infty}^\infty \frac 1{(z+ 2\pi n)^2}&=& \frac 14 {\rm cosec}^2\left(\frac z2\right),
 \eea
 then allow us to write  the evolution  equations in the form   
 \bea
i\dot a_j &=& \sum_{k;\,k \ne j} \frac{\lambda}{2}\cot \left(\frac{a_k-a_j}{2}\right) - \sum_{k=1}^M \frac{\lambda}{2}\cot \left(\frac {b_k-a_j}{2}\right),\
\label{EQ:pera_velocity}\\
i\dot b_j &=& \sum_{k;\, k \ne j} \frac{\lambda }{2}\cot \left(\frac{b_j-b_k}{2}\right) - \sum_{k=1}^N \frac{\lambda }{2}\cot \left(\frac{b_j-a_k}{2}\right),
\label{EQ:perb_velocity}
 \eea
 and 
  \be
 \ddot a_j = -\frac{\partial}{\partial a_j} \left\{\frac 14 \sum_{k;\, k \ne j} \frac{\lambda^2}{{\rm sin}^2(a_j-a_k)/2}\right\} .
 \label{EQ:sutherlandeq}
 \ee
 If we regard  the $a_i$   as angles,  then  (\ref{EQ:sutherlandeq})  
 is the equation of motion arising from  the Sutherland-model Lagrangian
 \be
 L_{\rm Sutherland}= \frac 12 \sum_{i=1}^{N} \dot a_i^2 - \frac 18 \sum_{i,j;\, i\ne j}\frac{\lambda^2}{{\rm sin}^2(a_i-a_j)/{2} }
 \label{EQ:sutherlandact}
 \ee
for  particles on a circle.

\begin{figure}
\includegraphics[width=3.5in]{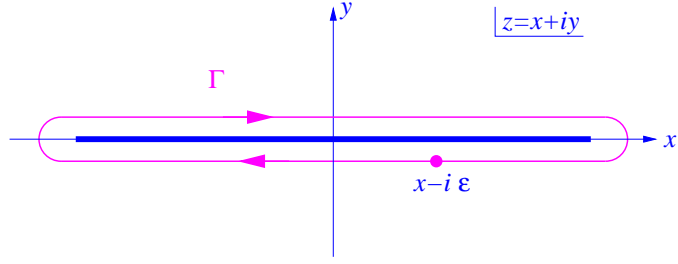}
\caption{\label{FIG:contour}{The contour $\Gamma$ surrounds the part of the real axis on which the $a_i$ poles are found.}}
\end{figure}

The authors of \cite{PLC}   made   $b_i=a_i^*$, as they  wished   $u(x,t)$ to be real.  Being interested primarily in  Calogero-Sutherland  models, we instead desire that the $a_i(t)$ be real.  To arrange for this, we  modify the definition of the Hilbert transform  (\ref{EQ:realHilbert}) appearing in 
 (\ref{EQ:boequation}).   Following Abanov and Wiegmann \cite{abanov}  we  define the $\Gamma$-contour Hilbert transform of $u$ to be
 \be
 u_\Gamma(z)\, \stackrel{\rm def}{=}\,\frac{\rm P}{\pi}\oint_\Gamma  \frac{1}{z-\xi}f(\xi)\,d\xi,
 \ee
 where $\Gamma$ is  a simple   closed   contour   on which $z$ lies. 
All  that was   needed to  establish   the $a$-pole autonomy  was  that  the  $1/(x-a_i)$'s and  $1/(x-b_i)$'s  be    eigenfunctions of the  Hilbert transform  with  eigenvalues of opposite sign.   
Now if we take  $\Gamma$ to  encircle  the  real axis in a clockwise sense (as shown  in Figure~\ref{FIG:contour})  then, for $z$ on the contour, 
 \bea
\left( \frac{1}{z-a_i}\right)_\Gamma &=& \frac{-i}{z-a_i},  \nonumber\\
\left( \frac{1}{z-b_i}\right)_\Gamma &=& \frac{+i}{z-b_i},                                                             
\eea
 when  the   $a_i$ poles  lie within $\Gamma$ and the $b_i$ poles lie outside.   
    We can therefore  let the $a_i(0)$  lie on the real axis  and  distribute  the  $b$-poles in the remainder of  the complex plane   in such a manner  that   the initial $\dot a_i$'s are real.  Once the intitial  $a_i$'s and $\dot a_i$'s are real, the Calogero evolution  ensures that the $a_i(t)$ remain on the real axis.  Thus, a complex Benjamin-Ono field $u(z,t)$  obeying 
 \be
\dot u +u\,\partial_z u= \textstyle{ 1\over 2}  \lambda  (\partial^2_{zz}u)_\Gamma
\label{EQ:complex_boequation}
\ee
on $\Gamma$ 
 can provide real-axis  Calogero-Sutherland dynamics.  

\section{Shepherd poles and  Calogero density-wave solitons}
\label{SEC:shepherd}

In this section we will apply the pole ansatz to obtain solitary wave  solutions to the hydrodynamic limit of   the real-axis Calogero equation.  In this limit  we let the number  $N$ of Calogero particles become infinite, while keeping their density finite.   We then  replace   the individual $a_i(t)$ by  a  smooth particle-number density  $\rho(x,t)$, and   the individual velocities $\dot a_i(t)$ by a smooth velocity field $v(x)$.  We assume that $\rho$ and $v$ vary slowly  on the scale of the inter-particle spacing.
We  are now   naturally tempted to approximate the  discrete $a$-pole sums in the evolution equations    
 \bea
 i\dot a_j&=& \sum_{k;\,k \ne j} \frac{\lambda}{a_k-a_j}- \sum_k \frac{\lambda}{b_k-a_j},\nonumber\\
 i\dot b_j&=&\sum_{k;\, k \ne j} \frac{\lambda}{b_j-b_k}- \sum_k \frac{\lambda}{b_j-a_k},
\eea
 by  integrals to get 
\bea
i\dot a(x,t) &=&{\rm P\!\!} \int_{-\infty}^{\infty} \frac{\lambda}{\xi-x}\rho(\xi,t)\,d\xi - \sum_k \frac{\lambda}{b_k-x},\nonumber\\
i\dot b_j(t)&=&\sum_{k;\, k\ne j} \frac{\lambda}{b_j-b_k } -\int_{-\infty}^{\infty} \frac{\lambda}{b_j-\xi} \rho(\xi,t)\,d\xi. 
\eea
Here, $\dot a(x,t)\equiv v(x,t)$ is the velocity of the pole at $x$.
We begin by  exploring  the consequences of this approximation to the discrete sum. We will see that it is not  quite  consistent, and a small but significant  correction is needed.

Consider  an  initial density fluctuation 
\be
\rho(x,0) = \rho_0+ \rho_1(x), \quad \hbox{where} \quad \rho_1(x)=\left(\frac{A}{\pi}\right)\frac{1}{x^2+A^2}.
\ee
Because
\be
\int_{-\infty}^{\infty} \left(\frac{A}{\pi}\right)\frac{1}{\xi^2+A^2}\, d\xi =1,
\ee
the Lorentzian  distribution  $\rho_1(x)$  corresponds to a local excess of  one   particle   near $x=0$. 

The contribution 
\be
{\rm P}\int_{-\infty}^{\infty} \frac{\lambda }{\xi-x}\left(\frac{A}{\pi}\right)\frac{1}{\xi^2+A^2}\,d\xi 
=   -\frac{\lambda x}{x^2+A^2}, \quad  ({\rm Im\,}x=0) 
 \ee
of the density fluctuation   to $i\dot a$ is real, and so  tends to push   $a(x)$ off the real axis. Its effect can be countered, however,   
 by placing a 
 solitary $b$-pole at at $b= iA$.  We then have 
\bea
i\dot a(x,0)&=& -\frac{\lambda x }{x^2+A^2} +\frac{\lambda}{x-iA}\nonumber\\
&=&  -\frac{\lambda x }{x^2+A^2} +\frac{\lambda(x+ iA)}{x^2+A^2}\nonumber\\
&=&   \frac{iA\lambda}{x^2+A^2}, 
\eea
and the  $a$ poles   have a purely  real initial velocity
\be
 v (x,0) {=} \dot a(x,0) = \frac{A\lambda}{x^2+A^2}.
\ee
They therefore stay on the real axis.

The motion of the $b=iA$ pole is  obtained from 
\bea
i\dot b& =& - \int_{-\infty}^{\infty} \frac{\lambda}{b-\xi} \left(\frac{A}{\pi}\right) \frac {1}{\xi^2+A^2}\, d\xi - \int_{-\infty}^{\infty} \frac{\lambda \rho_0}{b-\xi}\, d\xi\nonumber\\
&=& -\frac{\lambda}{b+iA} +i\lambda \pi\rho_0\nonumber\\
&=& \frac{i\lambda}{2A} +i\lambda \pi\rho_0.
\eea
Thus, the   $b$-pole  velocity
\be
\dot b= \frac{\lambda}{2A} +\lambda \pi \rho_0
\label{EQ:naive_pulsev}
\ee
is also purely real.
Since we know that the $a$ poles stay on the real axis,  it must be that the  $a$-pole density profile keeps  abreast of   the $b$ pole as it moves parallel to the real axis.  A  constant-shape  solitary wave  of density 
\be
\rho(x,t) = \rho_0+ \left(\frac{A}{\pi}\right)\frac{1}{(x-v_{\rm soliton}t)^2+A^2}
\ee
therefore moves to the right at the speed $v_{\rm soliton}= \dot b$.

We could  have  put the  $b$ pole   {\it below\/}  the axis. In that case both the $a$-poles and pulse envelope will move to the {\it left}.  In either case, the  envelope  velocity is always higher than the speed of sound $v_{\rm sound} =\lambda \pi \rho_0$,  and is faster when the pulse envelope is tighter.

Observe how the  $b$ pole acts as a  {\it shepherd\/}: its   real-part contribution serves to keep the $a$  poles from wandering off the real axis.  

The $a$ poles  are distributed  in such a manner  that the $b$ pole sees the   effect of their enhanced density as a mirror  image of itself lying below the axis.  The  contribution of this image  to the  $b$ velocity   parallels   the interaction of a  $b$ and $a=b^*$ pair in the original real-valued  Benjamin-Ono equation studied in \cite{PLC} --- the only difference being  that in the present case  we must  include in  $\dot b$ the constant velocity $v_{\rm sound}=\lambda \pi \rho_0$ induced by the uniform $\rho_0$ background.  
The condition that $\dot a(0)$  be real   is a  linear equation linking the $b$-pole contribution  to  the density fluctuation $\delta \rho=(\rho-\rho_0)$. Consequently  the  initial  data   for  a chiral $M$-soliton  can be established by placing  poles above the axis at $b_j$, $j=1,\ldots,M$. The associated  $\delta \rho$  is then the  sum of  individual Lorentzians centered at $x_j={\rm Re\,} b_j$, and   induces  image poles below the axis at $b_j^*$, $j=1,\ldots,M$.  Thus, in our present approximation, the   chiral Calogero hydrodynamics multisoliton   coincides with  the  multisoliton solution of the conventional Benjamin-Ono  equation. 
In particular, the real field  
\be
\widetilde u(x,t)\, \stackrel{\rm def}{=}\, \sum_{j=1}^M \left\{ \frac{i\lambda}{x-b_j^*(t)} -  \frac{i\lambda}{x-b_j(t)}\right\} + \lambda \pi \rho_0
\label{EQ:simple_u1}
\ee
obeys (\ref{EQ:boequation}), and possesses  the physical interpretations 
\be
\widetilde u(x,t)=v +\lambda\pi \rho= 2v  +v_{\rm sound}= \lambda \pi (2  \rho  -\rho_0).
\label{EQ:simple_u2}
\ee

This approximate mapping of the right-going Calogero density waves  onto the Benjamin-Ono solitons is very appealing.
Unfortunately there is a fly in the ointment:  the equation of continuity is  not  satisfied.    For  $v $ and $\rho$ both  being functions of of $x$ and $t$ in the combination $x-v_{\rm soliton} t$,  particle-number conservation  requires that 
\be
\dot  \rho+ \partial_x(\rho v )=  \partial_x\{\rho (v -v_{\rm soliton})\}=0.
\label{EQ:badwolf}
\ee
Because  $\rho\to \rho_0$ and $v  \to 0$ at large distance, the right-hand side  of
 (\ref{EQ:badwolf})   implies that 
\be
\rho (v_{\rm soliton}-v )= \rho_0 v_{\rm soliton},
\ee
or, equivalently,  
\be
v = \frac{\rho-\rho_0}{\rho} v_{\rm soliton}.
\label{EQ:soliton_continuity}
\ee
Now the  $a$-pole velocity and density that we have found  are linked by $v (x)= \lambda\pi (\rho(x)-\rho_0)$, and since $v_{\rm soliton}$ is fixed while   the $\rho$ in the numerator   varies with position,  
our present 
solution can obey (\ref{EQ:soliton_continuity}) only approximately.

\section{The non-linear  correction}
\label{SEC:correction}

To get the continuity equation to  hold exactly, we need to improve on  the crude    approximation 
\be
\sum_{k;\, k\ne j} \frac{\lambda}{a_k-a_j} \, \to \, {\rm  P}\int_{-\infty}^{\infty} \rho(\xi)\frac{\lambda}{\xi-a_j}\, d\xi. 
\ee
 This  na{\"\i}ve continuum approximation would be legitimate  (and  (\ref{EQ:soliton_continuity}) would hold exactly) were we to simultaneously let  the background density $\rho_0$ become infinite  and  take $\lambda\to 0$ in such a way that $\rho_0 \lambda$ remains constant. We are not doing this, however.   We  wish   to keep  $\lambda$ fixed while   $\rho_0$  becomes large,  but remains finite.  For finite $\lambda$,  the poles immediately adjacent to $a_j$  make a significant  contribution to the sum, and their effect  has to be carefully accounted for. 

An improved    approximation  to the pole sum is   
\be
\sum_{k;\,k\ne j} \frac{\lambda}{a_k-a_j}\,\sim\,  {\rm P\!\!}\int_{-\infty}^{\infty} \frac{\lambda}{\xi-a_j}\rho(\xi,t)\,d\xi  -\frac {\lambda}{2} \left.\partial_x \ln \rho(x)\right|_{x=a_j}.
\ee
The $\partial_x \ln \rho$ correction is the first  term in an asymptotic  series that expands   the  difference between the sum and integral  in    local gradients  of $\rho$.  It arises because  the particle at $a_j$ no longer  lies midway between its neighbours at $a_{j\pm1}$ when     the density varies with position --- but     the symmetric cutoff  in   the principal-part integral  tacitly assumes a midpoint location.  We provide a derivation of this  first  correction term  in the appendix.

Once we know of  the  local correction to the principal-part integral,  we realize that the shepherd  poles need to make a corresponding  additional    real  contribution    to $i\dot a$  if they are still to prevent   the $a_j$ poles from wandering off  the real axis. The equations determining this extra contribution are non-linear, and so the simple pole superposition that enabled us to find the multisolition initial conditions in the previous section is not longer valid.  We can still find some solutions, however.
As an illustration,  again consider  the initial density profile 
$$
\rho(x,0) =\rho_0 + \left(\frac{A}{\pi}\right) \frac {1}{x^2+A^2}.
$$
We now   have 
 have
\bea
{\rm P\!\!}\int_{-\infty}^{\infty} \frac{\lambda}{\xi-x}\rho(\xi,0)\,d\xi - \frac {\lambda}{2} \partial_x \ln \rho
&=& -\frac{\lambda x}{x^2+A^2} - \frac \lambda 2 \left\{ \frac A \pi \frac{-2x}{(x^2+A^2)^2}\frac{1}{\rho_0 + \displaystyle{\left(\frac{A}{\pi}\right) \frac{1} {x^2+A^2}}}\right\}\nonumber\\
&=& - \frac{\lambda x}{x^2+A^2+A/\pi \rho_0}\nonumber\\
&=& -\frac{\lambda x}{x^2+B^2},
\eea
where
\be
B=\sqrt{A^2+\frac{A}{\pi \rho_0}}.
\ee
We see that to  keep the $a$'s on the real axis, the  shepherd $b$-pole must   be relocated to  $b=iB$. The resultant $a$-pole motion is then  governed by  
\bea
i\dot a(x)&=& {\rm P\!\!}\int_{-\infty}^{\infty} \frac{\lambda}{\xi-x}\rho(\xi,0)\,d\xi - \frac {\lambda}{2} \partial_x \ln \rho+\frac{\lambda}{x-iB},\nonumber\\
&=& -\frac{\lambda x }{x^2+B^2} +\frac{\lambda}{x-iB},\nonumber\\
&=&  -\frac{\lambda x }{x^2+B^2} +\frac{\lambda(x+ iB)}{x^2+B^2},\nonumber\\
&=&   \frac{iB\lambda}{x^2+B^2}.
\eea
The new, improved, $v (x)$ is therefore 
\be
v (x) = \frac{\lambda B}{x^2+B^2}.
\label{EQ:vpole2}
\ee
From this, and (\ref{EQ:soliton_continuity}),  we get
\bea
v_{\rm soliton}&=& \frac{\rho}{\rho-\rho_0} v (x)\nonumber\\
&=& \rho_0 \lambda \pi\left( \frac BA\right), 
\label{EQ:vsoliton2}
\eea
which  is independent of $x$ as it should  be --- and was {\it not\/} previously. By squaring this last equation  we also  find  that 
\be
A\pi \rho_0 = \frac {v_{\rm sound}^2}{v_{\rm soliton}^2 -v_{\rm sound}^2},
\ee
which is the relation between  soliton width and velocity  obtained  in \cite{polychronakos_soliton}. 

Does this newly-computed soliton velocity  coincide with that of the shepherd $b$ pole?   
We set $b=iB$ in  
\be
i\dot b= -\frac{\lambda}{b+iA} +i\lambda \pi \rho_0,
\ee
to find  
\be
\dot b= \frac{\lambda}{A+B}+ \lambda\rho_0\pi.
\ee
Multiplying both sides by $A(A+B)$  shows that $\dot b=v_{\rm soliton}$, so the $b$ pole does indeed track the  density pulse --- as it  must  if the pulse is to retain its shape.  This consistency check also illustrates the fact  that the $\partial_x\ln \rho$ correction does not affect the contribution of the $a$-pole sum away from  the real axis. Once we are further away from the axis than the mean $a$-pole spacing, we no longer see the granularity of the $a$-pole ``charge'' distribution, and so the na{\"\i}ve integral suffices for calculating the $a$-pole contribution to the $b$-pole motion.  

We can also construct  periodic solutions. If we take these  to have period  $2\pi$,  the resulting wave trains   can be regarded as  solitons of the trigonometric Sutherland model.
Suppose, therefore, that the initial density profile is 
\bea
\rho(x,0) &=& \rho_0 +\sum_{n=-\infty}^{\infty} \left(\frac A\pi\right) \frac{1}{(x+2\pi n)^2 +A^2}\nonumber\\
&=& \rho_0 +\left( \frac 1{2\pi}\right) \frac{\sinh A}{\cosh A-\cos x}.
\eea
 Then 
\be
{\rm P\!\!}\int_{-\infty}^{\infty} \frac{\lambda}{\xi-x}\rho(\xi,0)\,d\xi  =   \frac{-(\lambda/2) \sin x}{\cosh A-\cos x},
\ee
and 
\be
{\rm P\!\!}\int_{-\infty}^{\infty} \frac{\lambda}{\xi-x}\rho(\xi,0)\,d\xi  - \frac \lambda 2 \partial_x \ln \rho
=  \frac{-(\lambda/2) \sin x}{\cosh A-\cos x +\displaystyle{\frac{ \sinh A}{2\pi \rho_0}}}.
\ee
This real contribution  to $i\dot a$ can be cancelled by placing  an infinite  train of shepherd poles at $z_n=iB+2\pi n$. From   
\be
\lim_{M\to \infty} \left\{\sum_{n=-M}^{M} \frac{1}{(x-iB)+2\pi n}\right\}= \frac 12  \cot \left(\frac {x-iB}{2}\right) = \frac 12 \left(\frac{\sin x+i\sinh B}{\cosh B-\cos x}\right),
\ee
we see that  $B$  should be chosen so that    
\be
\cosh B= \cosh A + \frac{1}{2\pi \rho_0} \sinh A.
\ee
By algebra  that parallels the single-soliton case,  we find that 
\be
v (x,0)= \left(\frac \lambda 2\right)  \frac {\sinh B}{\cosh B- \cos x},
\ee 
and that the soliton velocity is given by
\be
v_{\rm soliton} =\frac{\rho}{\rho-\rho_0}v (x)=  \rho_0\lambda \pi \left(\frac {\sinh B}{\sinh A}\right).
\ee

For a wave  train of  period $\Lambda$, these equations become
\be
v (x,0)= \left(\frac {\lambda \pi }{\Lambda} \right)
  \frac{\sinh (2 \pi B/ \Lambda)}{\cosh (2\pi B/\Lambda )- \cos(2\pi x/\Lambda)},
\ee 
and
\be
v_{\rm soliton} =  \rho_0\lambda \pi \frac {\sinh(2\pi B/\Lambda)}{\sinh(2\pi A/\Lambda)}.
\ee
They  reduce to (\ref{EQ:vpole2}) and (\ref{EQ:vsoliton2}), repectively,  when  $\Lambda$ becomes large.

\section{The general case}
\label{SEC:double}

We now show that the local  correction guarantees consistency with particle conservation for arbitrary initial data. 

To motivate  the discussion,  we begin by considering a simple model   
for the dynamics of a one-dimensional gas of   spinless, unit-mass, fermions.  The   uniform non-interacting gas would have internal energy density $\epsilon(\rho) = \hbar^2 \pi^2 \rho^3/6$.  To  include the effect of some  interactions, we  generalize this expression  by introducing a parameter $\lambda$ so that  $\epsilon(\rho) = \lambda^2   \pi^2 \rho^3/6$.  This $\lambda$ will later be identified with the parameter $\lambda$ appearing in the earlier sections. The simplest  classical Galiliean-invariant Hamiltonian for the gas is then
\be
H_{\rm fluid}= \int \left\{ \frac 12 \rho v^2 + \frac {\lambda^2 \pi^2}{6} \rho^3\right\}dx.
\ee
Here $v$ is the fluid velocity. We can write  $v=\partial_x \theta$,  where $\theta(x,t)$  is a phase field canonically conjugate to the density.  The field $\theta$   has Poisson bracket
\be
\{\theta(x),\rho(x')\} =\delta(x-x),
\ee
leading to
\be
\{v(x),\rho(x')\}= \partial_x \delta(x-x'), \quad \{v(x),v(x')\}=\{\rho(x),\rho(x')\}=0.
\label{EQ:fluid_poisson}
\ee
From $\dot \rho= \{H_{\rm fluid}, \rho\}$ we  obtain the equation of continuity
\be
\dot \rho+  \partial_x (\rho v)=0, 
\ee
and 
from  $\dot v= \{H_{\rm fluid},v\} $, we  obtain Euler's equation
\be
\dot v +v\partial_x v= - \partial_x\left( \frac {\lambda^2 \pi^2 \rho^2}{2}\right).
 \ee
 We can rearrange   $H_{\rm fluid}$ as 
\be 
H_{\rm fluid}= \frac 1 {2 \lambda \pi}  \int \left\{ \frac 1 6  (v+\lambda \pi \rho)^3 -\frac 16 (v-\lambda \pi \rho)^3 \right\}dx,
\ee 
and  the equations of motion  can similarly be massaged   into  Riemann form as 
\bea
\partial_t(v+\lambda\pi \rho) +(v+\lambda \pi \rho) \partial_x (v+\lambda \pi \rho)&=&0,\nonumber\\
\partial_t(v-\lambda\pi \rho) +(v-\lambda \pi \rho) \partial_x (v-\lambda \pi \rho)&=&0.
\eea
We therefore have   non-communicating  right-going and left-going Riemann invariants  
$I_{\rm R,L}(x) =  (v\pm \lambda \pi\rho)$  that are proportional to the chiral currents $j_{\rm R,L}= \frac 12(\rho\pm v/\lambda\pi)$ associated  with the right and left Fermi points. The Riemann  invariants   have  Poisson brackets 
\bea
\{I_{\rm R}(x),I_{\rm R}(x')\}\: &=& \phantom -{ 2\lambda \pi}\, \partial_x \delta(x-x'),\nonumber\\
\{I_{\rm L}(x),I_{\rm L}(x')\}\: &=& -{ 2\lambda \pi}\, \partial_x \delta(x-x'),\nonumber\\
\{I_{\rm R}(x),I_{\rm L}(x')\}\: &=&\phantom - 0.
\label{EQ:poisson2}
\eea 
Riemann's equations  also show that this simple  model  contains the seeds of its own destruction --- the leading edge of any simple wave will inevitably steepen and break, making the fields unphysically multivalued.

%\be
%H_{\rm CalogeroHa} =  \frac 12 \sum_{i=1}^{N} \dot x_i^2 + \frac 12 \sum_{i,j;\, i\ne j}\frac{\lambda^2}{(x_i-x_j)^2}.  
%\label{EQ:CalogeroHa}
%\ee

We now show  how the Calogero gas  mimics the Fermi gas,  while  regularizing  the multivaluedness.

Following \cite{abanov} we   decompose the $u$ field in  (\ref{EQ:poleansatz}) as  $u(z,t) =u_+(z,t)+u_-(z,t)$, where
 \bea
 u_-(z,t) &=&  \sum_{j=1}^N \frac{i\lambda}{z-a_j(t)},\\
u_+(z,t) &=& \sum_{j=1}^M  \frac{-i\lambda}{z-b_j(t)}.
\label{EQ:uplus}
 \eea
 The  $u_{\pm}(z,t)$ are eigenfunctions of the $\Gamma$-contour Hilbert transform with  eigenvalues $\pm i$ respectively. 
 In the hydrodynamic  limit,  $u_-$  becomes 
 \be
 u_-(z,t) =  i\int_{-\infty}^{\infty}   \rho(\xi,t)\frac{\lambda }{z-\xi}\,d\xi, 
  \ee 
 and has a discontinuity across the real axis:
 \be
 u_-(x+ i\epsilon)-u_-(x-i\epsilon)= 2\lambda \pi \rho(x).
 \ee
 Thus,
 \be
 u_-(x\pm i\epsilon)= \lambda \pi(i\rho_H\pm \rho).
 \ee
 With the local correction included,  the $a$-pole velocity  $v=\dot a$ is 
 \be
 v=  i\lambda \pi \rho_H + \frac {i\lambda} 2  \partial_x \ln \rho  +u_+.
 \ee
 We can rearrange the  last equation as 
  \be
 u_+(x,t) = v-i\lambda \pi \rho_H -\frac{i\lambda}{2} \partial_x \ln \rho.
 \label{EQ:uplusdef}
 \ee 
 For general   $\rho(x)$, $v(x)$, we must take (\ref{EQ:uplusdef})  to be  the {\it definition\/}  of $u_+(x)$ on the real axis.  We then   define $u_+(z)$  to be the analytic continuation of $u_+(x)$ away from  the axis. The analytically continued $u_+$  will have no singularities within $\Gamma$, but will  only be of the simple form (\ref{EQ:uplus}) for the    restricted  set   of initial data that leads to pure soliton solutions. Nonetheless, as we show in the appendix,   the only requirement for  $a$-pole autonomy   is that $(u_+)_\Gamma =i u_+$. 
 
The total  $u=u_++u_-$ field  therefore    has limits above and below  the real axis  equal to  
 \be
 u(x\pm i\epsilon) = v\pm \lambda \pi \rho -\frac{i\lambda}{2} \partial_x\ln \rho. 
 \label{EQ:ulimits}
 \ee
 These boundary values  $u(x\pm i\epsilon)$  almost coincide with the Riemann invariants   
 $I_{\rm R,L}(x)$.   The only difference is a shift  $v\to v- {i\lambda} \partial_x \ln \rho/2$, and this shift   does not affect  the Poisson algebra  in (\ref{EQ:poisson2}). In the quantum theory, the invariance comes about   because the shift is effected by  a conjugation \cite{stone_gutman}, and conjugation does not alter  c-number commutators.  In the classical theory, we   establish the invariance by observing that 
 \be
 \left\{\theta(x) - i\lambda\textstyle{\frac 12} \ln \rho(x), \theta(x') - i\lambda \textstyle{\frac 12} \ln \rho(x')\right\}=0,
 \ee
 and so, on differentiating by $x$ and $x'$, we find 
 \be
\left \{v(x)- i\lambda \textstyle{\frac 12}\partial_x  \ln \rho(x), v(x')- i\lambda\textstyle{\frac 12}  \partial_{x'} \ln\rho(x')\right\}=0.
 \ee
 The shifted velocity therefore still Poisson-commutes with itself. Showing the invariance of the other brackets is straightforward.
   In particular, we find 
 $u(x+ i\epsilon)$ has vanishing Poisson bracket with $u(x-  i\epsilon)$. Thus,  the real-axis  fluid-dynamics Poisson algebra (\ref{EQ:fluid_poisson})   is equivalent to  the natural extension  of the Benjamin-Ono Poisson algebra 
 (\ref{EQ:poisson1})  to the contour $\Gamma$. 
 
We can also reduce  the $\Gamma$-contour version of the Benjamin-Ono Hamiltonian
 \be
 H_\Gamma =\frac 1{2\lambda \pi}  \oint_\Gamma \left\{\frac 16 u^3 - \frac {\lambda} 4 u (\partial_z u)_\Gamma \right\}dz
 \label{EQ:gammabo}
 \ee
 to a real-axis integral. Using the expressions (\ref{EQ:ulimits}) for $u(x\pm i\epsilon)$, and $u_\Gamma=i(u_+-u_-)$   we  find, after some labour, that this  integral is   
 \be
 H_{\rm hydro} =  \int_{-\infty}^\infty \left\{ \frac 12 \rho v^2 + \frac {\lambda^2 \pi^2}{6} \rho^3- \frac{\lambda^2\pi}{2} \rho (\partial_x \rho)_H +\frac{\lambda^2}{8}   \frac{(\partial_x\rho)^2}{\rho}\right\}dx.  
  \label{EQ:CalogeroHb}
 \ee
 From the general theory, we know that (\ref{EQ:gammabo}) leads to the Benjamin-Ono equation  on $\Gamma$, and that that, in turn, leads to the Calogero  equation on the real axis. Thus 
 $H_{\rm hydro}$,  when used in conjunction with  the real-axis  fluid-dynamics Poisson algebra 
 (\ref{EQ:fluid_poisson}), must be the Hamiltonian governing the real axis hydrodynamics.  
 It  therefore  follows, from $\dot \rho=\{ H_{\rm hydro},\rho\}=-\partial_x(\rho v)$, that the equation of continuity is exactly satisfied. 
 
  $ H_{\rm hydro}$ is the continuum Hamiltonian of refs  \cite{polychronakos_soliton,jonke}.   In those works  it was  obtained  indirectly  by  taking $\hbar\to 0$  limit  of  the quantum continuum  Hamiltonian that was  already known from the collective field approach to matrix models \cite{andric0}.   In the appendix we provide a more direct, and purely classical, derivation of the rather less-than-obvious expression for  the  internal energy part of $H_{\rm hydro}$ :   
  \be
 V[\rho] =   \frac 12 \sum_{i\ne j} \frac {\lambda^2}{(a_i-a_j)^2}\, \sim\, \int_{-\infty}^\infty\left\{ \frac {\lambda^2 \pi^2}{6} \rho^3- \frac{\lambda^2 \pi}{2} \rho (\partial_x \rho)_H +\frac{\lambda^2}{8}   \frac{(\partial_x\rho)^2}{\rho}\right\}dx.
 \label{EQ:desired}
 \ee
 
 Although the boundary values $u(x\pm i\epsilon)$ of the $u(z,t)$ field Poisson commute, the  left- and right-going Riemann invariants  are no longer dynamically decoupled.   For example, an  inteplay  between the two segments  of the contour $\Gamma$ is clearly seen when we use the $u$-field boundary values to  compute the $\Gamma$-contour Hilbert transform of $u$ at $z=x-i\epsilon$:
 \bea
 u_\Gamma(x-i\epsilon)&=& \frac {\rm P}{\pi} \oint_\Gamma \frac {u(z')}{(x-i\epsilon)-z'}dz'\nonumber\\
 &=&\frac 1 \pi  \int_{-\infty}^\infty \frac{u(x'+i\epsilon)}{(x-i\epsilon)-(x'+i\epsilon)}dx' - \frac{\rm P}{\pi} 
 \int_{-\infty}^\infty \frac{u(x'-i\epsilon)}{x-x'}dx' \nonumber\\
 &=& \frac{\rm P}{\pi} \int_{-\infty}^\infty\frac{u(x'+i\epsilon)-u(x-i\epsilon)}{x-x'}dx' + \frac 1\pi\left\{i\pi u(x+i\epsilon)\right\}\nonumber\\
 &=& \frac{\rm P}{\pi} \int_{-\infty}^\infty \frac{2\pi \rho(x')}{x-x'}dx' +iu(x+i\epsilon),\nonumber\\
 &=& 2\pi \rho_H +i \left(v+ \lambda \pi \rho -{i\lambda}\textstyle{\frac 12} \partial_x\ln \rho\right)\nonumber\\
 &=& i\left.(u_+-u_-)\right|_{z=x-i\epsilon}.
 \eea
 In the second line, the two integrals are the contributions from the upper and lower segments  of $\Gamma$. The $u(x+i\epsilon)$ appearing in the third line is the delta-function contribution from
 \be
 \frac{1}{x-x'-i\epsilon}= {\rm P}\left(\frac {1}{x-x'}\right) +i\pi \delta(x-x')
 \ee
 in  the upper-segment integration.
 
 The interaction between the two branches of $\Gamma$ also shows up when we seek chiral ({\it i.e.}~unidirectional)  waves.  To have purely  right-going motion,  all the $u_+(z)$ singularities must  lie in the upper half-plane. From the theory of Hilbert transforms, this condition is equivalent to demanding that  
 \be
 (u_+)_H = iu_+.
 \label{EQ:chiral1}
 \ee
We can arrange for this eigenvalue condition  to hold by imposing the chiral constraint \cite{abanov,bettelheim,stone_gutman}
\be
v=\lambda \pi (\rho-\rho_0) -\frac 12 \lambda (\partial_x \ln \rho)_H,
\label{EQ:chiral2}
\ee
whence 
\be
u_+= \lambda \pi [(\rho-\rho_0)-i\rho_H] -\frac i2 \lambda\left[\partial_x\ln \rho-i  (\partial_x\ln \rho)_H\right].
\ee
This last form of $u_+$  obeys (\ref{EQ:chiral1}) because $(f_H)_H=-f$ for functions in 
$L^2({\mathbb R})$. (Note that  square-integrability  requires us to  subtract the background density from $\rho$ in these equations. The Hilbert transform of a constant is zero, and  $(\rho_H)_H=-(\rho-\rho_0)$.) 
Under the chiral condition, the equation of continuity and the Euler equation reduce to a   single  equation
\be
\dot \rho +\partial_x [\lambda\pi(\rho-\rho_0)\rho] = \frac{\lambda}{2} \partial_x[\rho(\partial_x \ln \rho)_H],
\ee
This strongly non-linear equation   reduces to the  right-going   Benjamin-Ono equation  for the  $\widetilde u=\lambda\pi(2\rho-\rho_0)$ field   introduced  in     equations  (\ref{EQ:simple_u1})  and 
(\ref{EQ:simple_u2}) if   we linearize  the dispersive term $\rho(\partial_x \ln \rho)_H \to  (\partial_x \rho)_H$, but  in  doing this   we abandon strict number conservation.

\section{Discussion}

We have seen  how the Benjamin-Ono equation on the ``double'' naturally  generates    the  solitary-wave and wave-train  solutions in   the classical hydrodynamic limit  of the Calogero-Sutherland model. We have also shown  how this approach  leads to  an understanding of  the the origin   the  subtle  corrections to the na{\"\i}ve continuous-fluid limit.  We could have computed these corrections to higher order in the gradients of $\rho$.  The small parameter in this  expansion is, however,    the local change in the inter-particle spacing divided by the spacing itself  and  should be  small in the hydrodynamic  limit.  Keeping track of these higher-order terms would  undo the advantages of the continuous-fluid  approximation. We therefore  retain only those that are required to maintain the internal consistency of the fluid mechanics model.  

It is interesting to compare the manner in which the classical machinery works with the   quantum mechanical formalism of \cite{abanov,stone_gutman}.  To make this comparison we should  first note that  the   parameter  $\lambda$  that appears in the quantum model is  a dimensionless number.  Our present classical parameter $\lambda$ has a scale, and is related to the quantum parameter by  
\be
\lambda_{\rm classical}= \hbar \lambda_{\rm quantum}. 
\ee
Thus, the free-fermion limit, where $\lambda_{\rm quantum}=1$,  corresponds to   $\lambda_{\rm classical}=\hbar$, and any  distinction between $\lambda_{\rm quantum}=1$ and $\lambda_{\rm quantum}=0$ is invisible in the classical $\hbar\to 0$ picture. What can be seen  in the classical picture is the analaytic structure of the $u(z,t)$ field and   the resulting interplay of  the left- and right-going currents $I_{\rm R,L}(x) \simeq u(x\pm i\epsilon)$ with    the positive and negative parts  $u_{\pm}(x)$ of the mode expansion.  In particular, we   see how these ingredients combine in a  Poisson-bracket algebra   that is the classical version  of the quantum mechanical  current algebra.

 \section{Acknowledgements}

We would like to thank Dmitry Gutman for stimulating our interest in these models, and for many discussions.  We also thank Larisa Jonke for comments on the manuscript.
This  work in was supported by the National Science Foundation under grant DMR-06-03528.

\section{Appendix}

In this appendix we provide proofs of some  assertions made in  the main text.

\subsection{Singular sums}

Here we obtain   the continuous-fluid  approximations to the sums 
\bea
S_1&=&\sum_{n=-\infty}^{\infty}  \frac{1}{a_n-a_0},\\
S_2 &=& \frac 12  \sum_{n,m=-\infty}^{\infty}   \frac{1}{(a_m-a_n)^2}.
\eea
  As usual,  the singular $a_n=a_0$ and $a_n=a_m$ terms  are to be omitted in their respective  sums. 
The essential tool is the Euler-Maclaurin expansion:
\be
\sum_{n=1}^\infty f(n) \sim \int_0^\infty f(\nu)\, d\nu  - \frac 12 f(0)  -\frac 1{12}f'(0) +\frac 1{720} f'''(0)+\cdots.
\ee 
This asymptotic expansion is valid for smooth  functions  $f(\nu)$.
The sums we are evaluating require  singular $f$'s, however, and so a strategy of subtractions is  needed.

We regard the discrete numbers  $a_n$ as the values   at $\nu =n\in {\mathbb Z}$  of  a   smooth monotonic-increasing function   $a(\nu)$. 
We wish to form the sum
\be
S_1=  \lim_{N\to \infty}\left\{\sum_{n=-N}^N\frac {1}{a(n) -a(0)}\right\},
\ee
where the term with $n=0$ is to be omitted. We replace this sum by   
\be
S'_1=\lim_{N\to \infty} \sum_{n=-N}^N \left\{ \frac {1}{a(n) -a(0)}- \frac{1}{a'(0)n}\right\},
\ee
where again the term with $n=0$ is to be omitted.  Now $S'_1=S_1$ because the   subtracted sum  is zero   ---  all its  terms cancel in pairs.
Thus 
\be   
 S_1 =  \lim_{N\to \infty}\left\{ \sum_{n=1}^N [f(n)+f(-n)]\right\},
 \ee
 where 
 \be
 f(\nu ) = \frac{1}{a(\nu)- a(0)}- \frac{1}{a'(0) \nu}
 \ee
 has a smooth $\nu \to 0$  limit: 
 \be
 \lim_{\nu\to 0} f(\nu)= -\frac 12 \frac{a''(0)}{[a'(0)]^2}.
 \ee
 Keeping only the first correction in the Euler-Maclaurin series, we have  
 \bea
 \int_0^\infty  f(\nu)\, d\nu&\sim & \frac 12 f(0) +f(1)+f(2)+\cdots,\nonumber\\
  \int_{-\infty}^0  f(\nu)\, d\nu&\sim & \frac 12 f(0) +f(-1)+f(-2)+\cdots,
 \eea
 and so 
 \be
 S_1 \sim  \int_{-\infty}^{\infty} f(\nu)\,d\nu +\frac 12 \frac{a''(0)}{[a'(0)]^2} +\cdots.
 \ee
 The integral in this last expression is convergent at $\nu=0$ since  $f(\nu)$ is finite there. Because the Hilbert transform of a constant vanishes, we can, however, remove the counter-term that makes $f(0)$  finite provided we  cut off the $\nu \to 0$ divergence with the principal-part prescription.  
 On doing  this, we obtain 
 \bea
 S_1&\sim & {\rm P} \int_{-\infty}^{\infty} \frac{1}{a(\nu)-a(0)} d\nu +\frac 12 \frac{a''(0)}{[a'(0)]^2}\nonumber\\
 &=&  {\rm P} \int_{-\infty}^{\infty} \frac{\rho(\xi)}{\xi-a_0} d\xi - \frac 12 \left. \partial_x \ln \rho(x)\right|_{x=a_0}
 \eea
 In passing from the first line to the second we have made a smooth change of variables $\nu \mapsto \xi=a(\nu)$  (such changes of variables are legitimate in principal-part integrals) 
 and defined the particle density $\rho(x)$ in terms of $a(\nu)$ by setting 
 \be
a'(\nu) = \frac{d a}{d  \nu}= \frac{d x }{d \nu} =  \frac 1{\rho(x)}.
 \ee
 We have also used  
 \be
 a''(\nu) = \frac{d^2 a}{d\nu^2}= -\frac 1{\rho^2} \frac{d \rho}{d\nu}= -\frac 1{\rho^3} \frac{d \rho}{dx }.
 \ee

To evaluate the double sum $S_2$, we begin by  obtaining the continuum approximation  to  
\be
S_3 = \sum_{n=-\infty}^{\infty} \frac 1{(a(n)-a(0))^2}
\ee
where the $n=0$ term is to be  omitted.
Following our subtraction strategy,  we use Euler's formula  $\sum_{n=1}^{\infty} 1/n^2 = \pi^2/6$ to write
\be
S_3 - \frac 1{[a'(0)]^2} \frac{\pi^2}{3} = \lim_{N\to \infty}\left\{ \sum_{n=1}^N [F(n)+F(-n)]\right\},
\ee
where 
\be
F(\nu)= \frac{1}{(a(\nu)-a(0))^2} - \frac 1{[a'(0)]^2}\frac{1} {\nu^2} + \frac{a''(0)}{[a'(0)]^3}\frac{1}{\nu}
\ee
has been constructed to have a finite $\nu\to 0$ limit:
\be
\lim_{\nu\to 0} F(\nu)= -\frac 13 \frac{a'''(0)}{[a'(0)]^3} +\frac 34 \frac{[a''(0]^2}{[a'(0)]^4}.
\ee
We apply the  Euler-Maclaurin formula as before to find  that 
\be 
S_3 \sim \frac{\pi^2}{3} \frac 1{[a'(0)]^2} +\int_{-\infty}^{\infty} F(\nu)\,d\nu -F(0) +\cdots.
\ee
We can again omit the the $1/\nu$ term in  $F(\nu)$ at the expense of replacing the convergent integral by a principal-part integral. 

We now claim that
\be
{\rm P} \int_{-\infty}^{\infty} \left\{ \frac 1{(a(\nu)-a(0))^2} - \frac 1 {[a'(0)]^2} \frac 1 {\nu^2} \right\}d\nu
=   {\rm P} \int_{-\infty}^{\infty} \frac {(\rho(x)-\rho(0))}{(x- a_0)^2}dx.
\ee
This result may be obtained by making a substitution  $\nu \mapsto x=a(\nu)$ in the first term of the integrand on the left-hand-side, and a substitution  $\nu \mapsto x= a(0)+  a'(0)\nu$ in the second term. Now it is not immediately obvious that the use of two distinct changes of variable is legitimate. The integrals of the two terms do not exist separately -- only the integral of their difference   is convergent at $\nu=0$.  In order to be sure that no additional finite contribution in introduced by our  man{\oe}uvre, we must provide a common $|\nu|>\epsilon $ cutoff for   the two separate integrals,  and then keep track of the effect of the  subsequent changes of variables on their  integration limits.  Because the two changes of variables agree to linear order near $\nu=0$, we find   that no such finite additions are induced. Our  man{\oe}uvre is indeed allowed.
We also note that
\be
 {\rm P} \int_{-\infty}^{\infty} \frac {(\rho(\xi )-\rho(0))}{(\xi-x)^2}d\xi = \frac{d}{dx} \left({\rm P} \int_{-\infty}^{\infty} \frac {\rho(\xi)}{\xi-x} d\xi \right)= -\pi \partial_x( \rho(x)_H) =-\pi (\partial_x \rho)_H. 
\ee
Next we  re-express $F(0)$ as 
\be
F(0)= - \frac 14 \frac{[a''(0)]^2}{[a'(0)]^4} - \frac 13 \frac{d}{d \nu} \left.\left(\frac{a''(\nu)}{[a'(\nu)]^3}\right)\right|_{\nu=0}.
\ee
To complete the   evaluation  of $S_2$  we should replace $a_0$ by $a_n$ and sum over $n$. In the continuum aproximation we replace  $a(0)$ by $a(\nu)$ and integrate over $\nu$. The  total derivative in $F$ will not contribute to this integral and can be discarded.  After changing variables $\nu\to x$, we   therefore find that
\be
S_2 \sim \int_{-\infty}^{\infty} \left\{ \frac{\pi^2}{6} \rho^3 - \frac{\pi}{2} \rho (\partial_x \rho)_H +\frac 18 \frac{(\partial_x  \rho)^2}{\rho} \right\}dx +\cdots,
\ee
which immediately gives  equation (\ref{EQ:desired}).

\subsection{Pole-autonomy for a general $u_+$.}

We here  show that  we can relax the condition that $u_+$ be a sum of simple poles, yet  still have the 
$a_j$ obey the Calogero equation (\ref{EQ:calogeroeq}).

Let
\be
u_-(z,t)= \sum_{j=1}^N \frac{i\lambda}{z-a_j(t)},
\ee
as before, but   assume  of $u_+(z,t)$ only that $(u_+)_\Gamma=+i u_+$.
Insert $u=u_-+u_+$ into
\be
\dot u + u \partial_z u = \textstyle{\frac 12} \partial_{zz}^2 u_\Gamma.
\label{EQ:appendixBO}
\ee
 The projections of the product $u_+u_-$ onto  the $\pm i$ eigenspaces of the Hilbert transform are  respectively 
\bea
(u_+u_-)_+&=& u_+u_- - \sum_{j=1}^N \frac{i\lambda}{z-a_j} u_+(a_j),\nonumber \\
(u_+u_-)_-&=&   \sum_{j=1}^N \frac{i\lambda}{z-a_j} u_+(a_j).
\label{EQ:pmprojections}
\eea
By taking the $z$ derivative of (\ref{EQ:pmprojections}), we 
we can  project  the cross terms $u_+\partial_zu_-+u_-\partial_z u_+$  appearing   in (\ref{EQ:appendixBO})   into the $\pm i$ eigenspaces .  From the coefficients of $1/(z-a_j)^2$ in  the $-i$ eigenspace, we   read off  from (\ref{EQ:appendixBO})  that
\be
i\dot a_j =\sum_{k;\, k\ne j} \frac{\lambda}{a_k-a_j} +iu_+(a_j).
\ee
In  the  $+i$ eigenspace, we  find that  (\ref{EQ:appendixBO})  requires that
\be
\dot u_+ +u_+\partial_z u_+ + \sum_{k=1}^N \frac{i\lambda}{z-a_k} + \sum_{k=1}^N \frac{i\lambda}{(z-a_k)^2}\left(u_+(a_k)-u_+(z)\right)=i\lambda \textstyle{\frac 12} \partial_{zz}^2 u_+.
\label{EQ:uplus_evolution}
\ee

Now differentiate $\dot a_j$ again to find that
\be
i\ddot a_j = \sum_{k;\, k\ne j}\frac{\lambda}{(a_k-a_j)^2}(\dot a_j-\dot a_k) + i\left(\dot u_+(a_j) +\dot a_j \partial_z u_+|_{z=a_j}\right).
\ee
If we momentarily forget all about $u_+$, we know that we can assemble the remaining terms to obtain the Calogero equation
$$
\ddot a_j = \sum_{k;\, k\ne j}\frac{2\lambda^2}{(a_j-a_k)^3}.
$$
We therefore need to show that all terms in $i\ddot a_j$ involving  $u_+$ drop out. These terms are
\be
\sum_{k;\,k\ne j} \frac{\lambda}{(a_k-a_j)^2} \left(u_+(a_j)-u_+(a_k)\right) +
i\left(\dot u_+(a_j)+\sum_{k;\,k\ne j} \frac{i\lambda}{a_j-a_k} \partial_zu_+|_{a_j} +u_+\partial_zu_+|_{a_j}\right).
\label{EQ:desire_zero}
\ee
 
Now consider the limit of  (\ref{EQ:uplus_evolution})  as $z\to a_j$. There are potential singularities arsising from the $k=j$ terms in the sums, but on  expanding
$$
u_+(z)=u_+(a_j) +(z-a_j)\partial_zu_+|_{a_j} +\textstyle{\frac 12}(z-a_j)^2 \partial_{zz}^2u_+|_{a_j}+O\left[(z-a_j)^3\right], 
$$
 we find that the potentially-singular parts  cancel among themselves, and, futhermore, the remaining   finite parts of the $j=k$ terms  combine to cancel   the   $i\lambda \textstyle{\frac 12} \partial_{zz}^2 u_+$ term on the right hand side. The $z=a_j$ limit of   (\ref{EQ:uplus_evolution}) thus  equates to zero  precisely  the terms (\ref{EQ:desire_zero}) that we wish to disappear.

\end{document}